\documentclass[twocolumn,secnumarabic,amssymb, nobibnotes, aps, prb,groupedaddress,superscriptaddress]{revtex4-2}

\usepackage{amsmath,graphicx,latexsym,times,color}
\usepackage{setspace}
\usepackage{hyperref}
\usepackage{array}
\usepackage{textcomp}
\usepackage{titlesec}
\usepackage{physics}
\usepackage{gensymb}
\usepackage{nccmath}
\usepackage{empheq} 
\usepackage{xcolor}

\definecolor{blue-violet}{rgb}{0.54, 0.17, 0.89}\newcommand{\V}[1]{\ensuremath{\mathbf{#1}}} 

\let\oldtimes\times  
\renewcommand\times{{\oldtimes}}

\begin{document}
	

\title{Long lifetimes of nanoscale skyrmions in lithium-decorated van der Waals ferromagnet Fe$_3$GeTe$_2$}
 
\author{Soumyajyoti Haldar}
\affiliation{Institute of Theoretical Physics and Astrophysics, University of Kiel, Leibnizstrasse 15, 24098 Kiel, Germany}

\author{Moritz A. Goerzen}
\affiliation{CEMES, Universit\'e de Toulouse, CNRS, 29 rue Jeanne Marvig, F-31055 Toulouse, France}
	
\author{Stefan Heinze}
\affiliation{Institute of Theoretical Physics and Astrophysics, University of Kiel, Leibnizstrasse 15, 24098 Kiel, Germany}
\affiliation{Kiel Nano, Surface, and Interface Science (KiNSIS), University of Kiel, 24118 Kiel, Germany}

\author{Dongzhe Li}
\email[Contact author: ]{dongzhe.li@cemes.fr}
\affiliation{CEMES, Universit\'e de Toulouse, CNRS, 29 rue Jeanne Marvig, F-31055 Toulouse, France}

\date{\today}
	
\begin{abstract}
 
The Dzyaloshinskii-Moriya interaction (DMI), which originates from spin-orbit coupling and relies on broken inversion symmetry, is recognized as a key ingredient in forming magnetic skyrmions. However, most 2D magnets exhibit inversion symmetry; therefore, the DMI is suppressed. Here, we propose a strategy to induce large DMI via lithium absorption on the surface of 2D magnets -- an experimentally feasible approach. Using first-principles and atomistic spin simulations, we predict the formation of nanoscale skyrmions in lithium-decorated monolayer Fe$_3$GeTe$_2$ by imposing small out-of-plane magnetic fields ($B_z$). Notably, we find very large skyrmion energy barriers of more than 300 meV at $B_z = 0.4$ T, comparable to those observed in ferromagnet/heavy-metal interfaces. The origin of these unique skyrmions is attributed to the competition between strong DMI, exchange frustration, and small magnetocrystalline anisotropy energy. We further show that the lifetimes of metastable skyrmions exceed one hour for temperatures up to 75 K. 
\end{abstract}
	
\maketitle

Magnetic skyrmions -- chiral, localized spin textures protected by an integer topological charge -- are prime candidates for the next generation of spintronic devices \cite{fert2017magnetic,everschor2018perspective,Luo2018,gobel2021beyond,li2021magnetic,Psaroudaki2021}. The key ingredient in forming magnetic skyrmions is the so-called Dzyaloshinskii-Moriya interaction (DMI) which arises in magnetic materials with intrinsic broken inversion symmetry, combined with strong spin-orbit coupling (SOC). Therefore, in the last decade, the main focus of the community has been devoted to constructing 
ferromagnet/heavy-metal interfaces with perpendicular magnetic anisotropy and large DMI to generate skyrmions \cite{heinze2011spontaneous,Romming2013,dupe2014tailoring,moreau2016additive,boulle2016room,soumyanarayanan_2017,meyer2019isolated}.

Between 2016 and 2017, the first experimental confirmation of magnetism in atomically thin 2D materials was reported \cite{Lee2016,gong2017discovery,huang2017layer}, opening up new opportunities for exploring novel magnetic phenomena in reduced dimensions. In 2020, magnetic skyrmions were first experimentally observed in 2D van der Waals (vdW) magnets \cite{han2019topological,ding2019observation}, providing an ideal playground to advance skyrmion technology towards the single-layer limit with large tunability via external stimuli. However, the DMI in most 2D magnets is suppressed due to their intrinsic inversion symmetry. Several strategies have been proposed to obtain broken inversion symmetry. The family of monolayer Janus vdW magnets, which lacks inversion asymmetry with different atoms occupying top and bottom layers, has been predicted theoretically to induce large DMI to generate skyrmions \cite{Liang2020,Changsong2020,du2022spontaneous}. However, only a few of these magnetic Janus monolayers have been experimentally realized \cite{xu2025unusual, nie2024regulated}.  
The electric field was also used to obtain broken inversion symmetry in 2D magnets to generate sizable DMI \cite{Liu2018}. Additionally, constructing various 2D vdW heterostructures have been proposed to tune and obtain DMI \cite{wu2020neel,wu2021van,Dongzhe2022_fgt}. In particular, Néel-type magnetic skyrmions are reported in Fe$_3$GeTe$_2$ heterostructures by experiments \cite{wu2020neel,yang2020creation,wu2021van} and explained by \textit{ab initio} theory \cite{Dongzhe2022_fgt,Dongzhe_prb2023} in terms of the emergence of strong DMI at interfaces. All-electrical skyrmion detection has also recently been proposed in tunnel junctions based on Fe$_3$GeTe$_2$ \cite{li2023proposal}. More recently, decorating light atoms on the surface of 2D magnets turns out to be an effective way to obtain sizable DMI and finally stabilize individual magnetic skyrmions \cite{Park2021,chen2020large,Cheng_PRB2023,thomsen2023magnetic,Weiyi2024}.

\begin{figure}[t]
\centering
\includegraphics[width=1.0\linewidth]{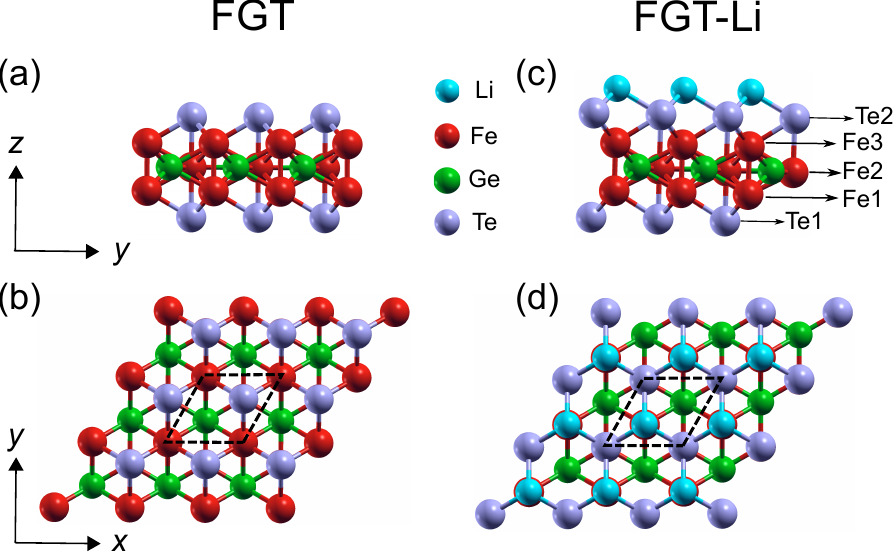}\\
\caption{\label{FGT_structure} (a) Side and (b) top views of the atomic structure of the Fe$_3$GeTe$_2$ monolayer. (c-d) The same as (a-b) but for lithium-decorated monolayer Fe$_3$GeTe$_2$. The black dashed lines denote the 2D primitive unit cell.}
\end{figure}
In this Letter, we propose a strategy to induce large DMI by decorating Li atoms on the surface of 2D magnets, which have been widely used to modulate the properties of 2D layered systems \cite{Eunseok2012,Chilin2012,XuYong_PRL2020}. We demonstrate the idea using monolayer Fe$_3$GeTe$_2$, a material currently generating significant experimental interest \cite{ding2019observation, Park2021,wu2020neel,wu2021van,powalla2023seeding}. Using first-principles calculations, we obtain
a large DMI and a significantly suppressed perpendicular magnetocrystalline anisotropy energy (MAE) in Li-decorated monolayer Fe$_3$GeTe$_2$ (FGT-Li) due to strong hybridization effect between the Li atom and the FGT layer. Our atomistic spin simulations (with the magnetic interaction parameters from first-principles) predict the formation of a spin spiral state in the absence of a magnetic field. Upon applying small out-of-plane magnetic fields ($B_z$), the nanoscale skyrmion phase is formed, which persists within a rather wide $B_z$ range. In particular, we predict very large skyrmion energy barriers of more than 300 meV at $B_z = 0.4$ T, which is among the highest reported for 2D magnets. Furthermore, we show that the lifetimes of metastable skyrmions in the ferromagnetic ground state are above one hour for temperatures up to 75 K.

\begin{figure}[t]
\centering
\includegraphics[width=0.96\linewidth]{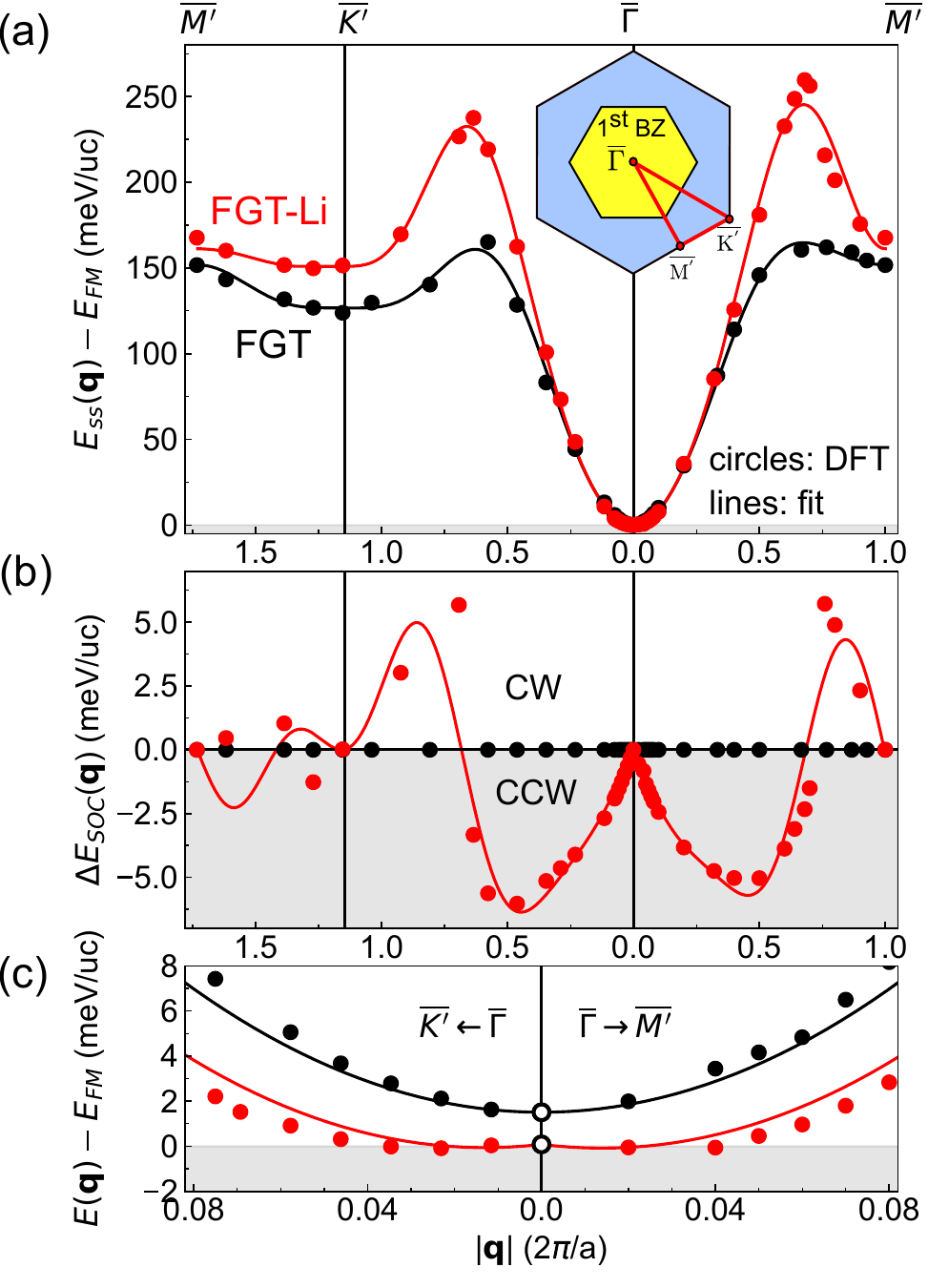}
\caption{\label{spin_spiral} (a) Energy dispersion of flat cycloidal spin spirals ($E_{\text{SS}}$) for FGT (black) and FGT-Li (red) along the high symmetry path $\overline{\text{M}^{\prime}\text{K}^{\prime}{\Gamma}\text{M}^{\prime}}$ without SOC. The symbols represent the DFT calculations, while the solid lines are the ﬁts to the Heisenberg model. A sketch of the 2D hexagonal Brillouin zone beyond the first BZ (yellow part), including high-symmetry points, is shown as an inset. (b) Energy contribution of cycloidal spin spirals due to SOC ($\Delta E_{\text{SOC}}$). All energies are measured with respect to the FM state at the $\overline{\Gamma}$ point. Note that positive and negative energies represent a preference for clockwise (CW) and counterclockwise (CCW) spin configurations. (c) Zoom around the $\overline{\Gamma}$ point, including the Heisenberg exchange, the DMI, and the MAE, i.e.~$E(\V{q})=E_{\rm SS}(\V{q})+\Delta E_{\text{SOC}}(\V{q})+K/2$. The DMI leads to CCW chirality,
and the MAE is responsible for the constant energy shift ($K/2$) of the spin spirals with respect to FM.}
\end{figure}
To determine the interactions of magnetic moments for the free-standing FGT monolayer and the FGT-Li monolayer, we have performed first-principles total energy calculations using the \textsc{Fleur} code \cite{fleurv26} and mapped these onto the following atomistic spin Hamiltonian:

\begin{equation}\label{model}
\begin{split}
H & =-\sum_{ij}J_{ij}(\V{m}_i \cdot \V{m}_j)-\sum_{ij}\V{D}_{ij} \cdot(\V{m}_i \times \V{m}_j) \\
& +\sum_i K_i (m_i^z)^2 - \sum_i \mu (\V{m}_i \cdot B_z)
\end{split}
\end{equation}
where $\V{m}_i$ and $\V{m}_j$ are normalized magnetic moments at position $\V{R}_i$ and $\V{R}_i$ respectively. The four magnetic interaction terms correspond to the Heisenberg isotropic exchange, the DMI, the magnetocrystalline anisotropy energy (MAE), and the external magnetic field, and they are characterized by the interaction constants $J_{ij}$, $\V{D}_{ij}$, and $K_i$, and $B_{\text{ext}}$, respectively. 

In order to determine the exchange interaction for arbitrary nearest neighbors, we used the energy dispersions of flat spin spirals, $E_{\text{ss}} (\V{q})$, calculated via DFT without SOC based on the generalized Bloch theorem \cite{Kurz2004}. Then, the SOC contribution to the energy dispersion of spin spirals, $\Delta E_{\text{SOC}} (\V{q})$, was calculated in first-order perturbation theory \cite{Heide2009,Zimmermann2014}. By fitting this energy contribution to the spin model, the DMI for arbitrary nearest neighbors was determined. Computational details are given in Section S1 in the Supplemental Material (SM) \cite{supplmat}. 

FGT exhibits the space group (194) P6$_3$/\textit{mmc} as shown in Fig. \ref{FGT_structure}(a). Subsequently, the top, center, and bottom atoms are denoted as Fe3, Fe2, and Fe1, respectively. For freestanding FGT, the DMI involving either Fe1 or Fe3 has opposite signs (i.e., chirality) because of the (001) mirror plane, resulting in zero total DMI. Upon absorption Li atom on the surface of monolayer FGT (denoted as FGT-Li in the following, see Fig. \ref{FGT_structure}(b)), the centrosymmetry of the system is broken, giving rise to sizable DMI. The energetically most favorable configuration involves the Li atom occupying a hollow site with 3 bonds with Te atoms, specifically positioned atop the Fe3 atom. The distance between the Li and Te atom is about 2.64 \AA, while it is about 2.79 \AA~for the Li-Fe3 distance. The calculated absorption energy is found to be about $-1.52$ eV, indicating a strong hybridization effect. Both freestanding FGT and FGT-Li exhibit a metallic property (See Fig.~S2 and Table S3 in SM~\cite{supplmat}). The calculated spin moments are $2.44\mu_{\text{B}}$ ($2.28\mu_{\text{B}}$), $1.58\mu_{\text{B}}$ ($1.40\mu_{\text{B}}$), $2.18\mu_{\text{B}}$ ($2.28\mu_{\text{B}}$), and $0.01\mu_{\text{B}}$ ($0.00\mu_{\text{B}}$) for the Fe1, Fe2, Fe3, and Li atoms of FGT-Li (freestanding FGT), respectively. We also checked the stability of the FGT-Li monolayer. As shown in Fig. S1 in SM \cite{supplmat}, the phonon spectrum shows only tiny negative frequencies throughout the Brillouin zone (BZ), indicating its dynamical stability.

\begin{figure}[htbp]
	\centering
	\includegraphics[width=0.9\linewidth]{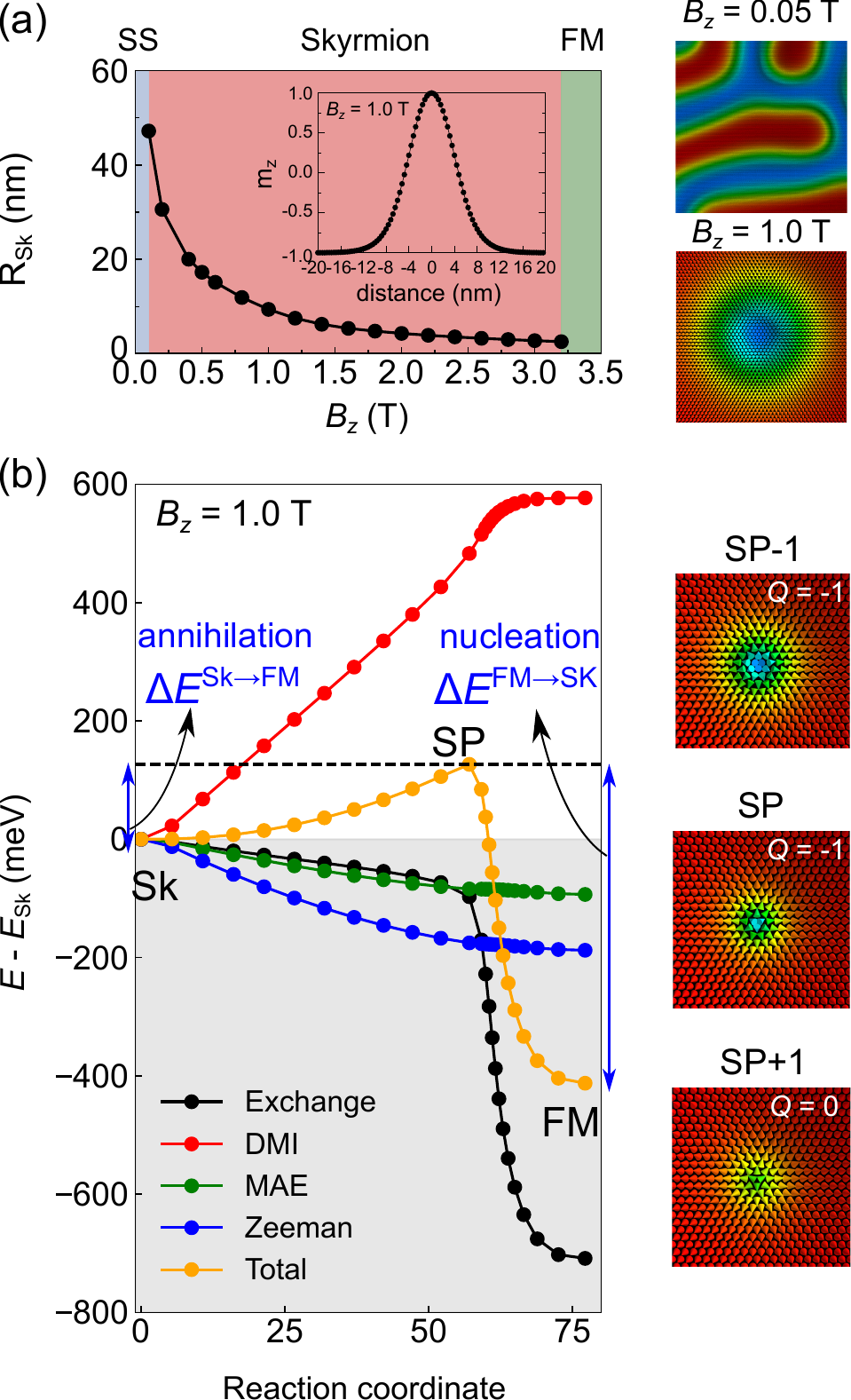}
	\caption{\label{GNEB} Skyrmion radius and MEP for skyrmion collapse in FGT-Li evaluated using magnetic interactions from DFT. (a) Skyrmion radius as a function of the applied magnetic field ($B_z$). Individual skyrmions can be formed in a broad range of $B_z$ from 0.1 $-$ 3.2 T (plotted as a red area), in which $B_z^{\text{c}} = 0.1$ T is the critical magnetic field. The range below $B_z^{\text{c}}$ is plotted as a light blue area.
    The skyrmion profile at $B_z = 1.0$ T is shown as an inset. Spin structures at $B_z = 0.05$ T (spin spiral state) and $B_z = 1.0$ T (Néel-type skyrmion) are shown on the right side. (b) MEP for a transition between the skyrmion (Sk) and the FM state at $B_z = 1.0$ T.  The energy contributions of the different interactions are represented by the color code (see legend). The saddle point of the transition is marked and the activation energies for skyrmion nucleation $\Delta E ^{\text{FM} \rightarrow \text{Sk}}$, and skyrmion annihilation $\Delta E ^{\text{Sk} \rightarrow \text{FM}}$, are indicated. Corresponding spin structures before (SP-1), after (SP+1), and at the saddle point (SP) with corresponding topological charges ($Q$) are shown. The skyrmion collapse occurs via radially symmetrical shrinking.}
\end{figure}

We focus first on spin spiral calculations without SOC (Fig.~\ref{spin_spiral}(a)) for a spin spiral vector $\mathbf{q}$ along the high symmetry paths ${\overline{\Gamma\mathrm{M}'}}$ and ${\overline{\Gamma\mathrm{K}'\mathrm{M}'}}$ of its BZ (beyond the first BZ). This is needed in order to describe accurately the magnetic interactions in the complex geometrical structure of FGT, for details, please refer to Ref.~\cite{Dongzhe_bimeron2024}. The high-symmetry points represent special states: the $\overline{\Gamma}$ point corresponds to the ferromagnetic (FM) state, the $\overline{\text{K}^{\prime}}$ point to the N\'eel-state with 120$^{\circ}$ between adjacent spins, and the $\overline{\text{M}^{\prime}}$ point to the row-wise antiferromagnetic (AFM) state (Fig.~\ref{spin_spiral}(a) inset). Here, to efficiently describe the three Fe atoms in each unit cell of FGT and due to the strong exchange coupling between neighboring Fe atoms, we treat them within the spin model
as one magnetic atom in a hexagonal lattice 
as in our previous work \cite{Dongzhe_prb2023}. Additionally, the use of a large $\V{q}$ vector beyond the first BZ is necessary due to the honeycomb nature of FGT (see SM~\cite{supplmat} for more technical details). A fit of the energy dispersion to the Heisenberg model yields the energy parameters as shown in Table S1 in SM \cite{supplmat}. All magnetic interaction parameters are measured in meV/unit cell (uc). In the following, we mainly focus on the comparison between spin spiral curves for FGT (black) and FGT-Li (red) in order to understand the influence of the Li atom on the magnetic interactions for the FGT layer. 

Without SOC, for both FGT and FGT-Li, the energy dispersion shows a minimum at the $\overline{\Gamma}$ point, which represents the FM state (Fig.~\ref{spin_spiral}(a)). The dispersion curve for FGT-Li (red) has a much larger energy scale compared to the one for FGT, indicating the significantly enhanced Heisenberg exchange in FGT-Li (also reflected in the fitted parameters in Table S2). When SOC is turned on, the DMI in the FGT monolayer is zero due to inversion symmetry, as expected, while the DMI arises for FGT-Li due to broken inversion symmetry (Fig.~\ref{spin_spiral}(b)). The DMI energy strength observed in FGT-Li is comparable to that in  strained FGT/Ge \cite{Dongzhe2022_fgt}. The FGT-Li monolayer favors a counter-clockwise (CCW) rotational sense, as indicated by the 
negative energy contributions due to SOC, $E_{\text{DMI}}(\V{q})$, to the dispersion of cycloidal spin spirals (Fig.~\ref{spin_spiral}(b)). 
Our DFT calculations indicate that the DMI mainly originates from interfacial nonmagnetic atoms (see Fig.~S2 in SM \cite{supplmat}). This mechanism is the same as the one reported in strained FGT/Ge \cite{Dongzhe2022_fgt} (see Fig.~S3 in SM \cite{supplmat}).
In order to understand the electronic effect of Li decoration in DMI and distinguish its origin from structural relaxation effect, we have calculated the energy contributions due to SOC (Fig.~S5 in SM~\cite{supplmat}) for Li decorated unrelaxed FGT system. Our calculation shows that $\Delta E_{\textrm{soc}}$ is $\sim$ 4 meV/uc for the Li-decorated unrelaxed FGT system (Fig.~S5). 
Therefore, the large DMI originates from the broken inversion symmetry due to Li decoration as understandable within the Fert-L\'{e}vy model~\cite{Fert1980} in combination with
the structural relaxation of FGT layers due to the Li adsorption.
Additionally, a significant reduction in the out-of-plane MAE is observed, shifting from $-1.0$~meV/Fe to $-0.05$~meV/Fe upon the Li-absorption onto the surface of monolayer FGT.

The total spin spiral energy dispersion, $E(\mathbf{q})$, includes all magnetic interactions, i.e.~exchange, DMI, and MAE (Fig.~\ref{spin_spiral}(c)). The energy contribution from the MAE leads to an energy offset of $K/2$ for spin spirals with respect to the FM state, as can be seen in a zoom of $E(\V{q})$ around $\overline{\Gamma}$. We find that the ground state for FGT is the FM state. In contrast, it is a spin spiral state for FGT-Li 
since there is an energy minimum in $E(\mathbf{q})$ of about $-$0.07 meV/uc compared to FM (Fig.~\ref{spin_spiral}(c)) which occurs in the $\overline{\Gamma\text{K}^{'}}$ direction, and corresponds to a spin spiral period of $\lambda = 2\pi/\V{|q|} \approx 57$~nm. It is worth emphasizing that the spin spiral curve becomes extremely flat near the $\overline{\Gamma}$ point for FGT-Li. It has been demonstrated that such an energy dispersion is beneficial for stabilizing nanoscale skyrmions in ultrathin films \cite{meyer2019isolated}.

\begin{figure*}[htbp]
\centering
\includegraphics[width=1.0\textwidth]{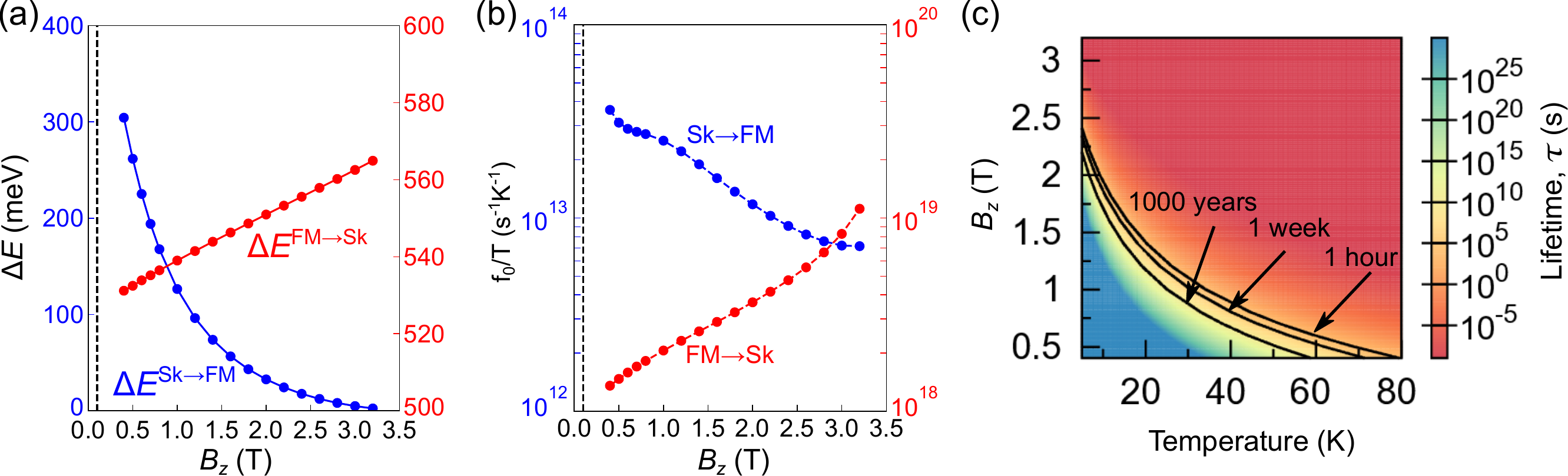}
\caption{\label{Lifetime} Skyrmion energy barrier, pre-exponential factor, and lifetime for FGT-Li. (a) Skyrmion collapse ($\Delta E ^{\text{Sk} \rightarrow \text{FM}}$, blue) and creation barrier ($\Delta E ^{\text{FM} \rightarrow \text{Sk}}$, red) as a function of $B_z$. (b) Calculated prefactors divided by temperature, $f_0$/T, are shown on a logarithmic scale. Blue and red curves denote $f_0$/T for skyrmion collapse and creation transitions, respectively. The vertical black dashed line represents the critical magnetic field $B_z^\text{c}$ = 0.1 T for obtaining magnetic skyrmions. (c) The lifetime of skyrmions, $\tau$, in FGT-Li obtained in harmonic transition state theory based on the spin model with DFT parameters.}
\end{figure*}
To explore the possible topological spin textures in monolayer FGT-Li, we performed atomistic spin simulations with ﬁrst-principles parametrized Hamiltonian of Eq.~(\ref{model}). In order to obtain isolated magnetic skyrmions that correspond to local energy minima with respect to the Hamiltonian in Eq. (\ref{model}), we employed the velocity projection optimization algorithm (VPO) \cite{bessarab2015method}. At low out-of-plane magnetic fields of $B_z = 0.05$~T, labyrinth domains are observed (Fig.~\ref{GNEB}(a)). When we increase $B_z$ up to the critical magnetic field $B_z^\text{c}$ = 0.1 T, the labyrinth domains disappear and isolated skyrmions (Néel-type) emerge in the FM background as seen from spin structures at $B_z= 0.05$~T and 1.0~T in Fig.~\ref{GNEB}(a). Importantly, the skyrmion phase is preserved within a wide field range of $0.1 \sim 3.2$ T. 
As expected, we observe that the skyrmion size decreases with increasing ﬁeld. Here, the skyrmion radius is estimated from the relaxed skyrmion profiles (see inset of Fig.~\ref{GNEB}(a)) using the definition of Bogdanov \textit{et al.} \cite{bocdanov1994properties}. By further increasing the magnetic field, the density of skyrmions decreases and drops to zero at 3.4~T, which gives rise to the 
field-polarized FM phase.

We used the geodesic nudged elastic band (GNEB) method ~\cite{bessarab2015method} to calculate the minimum energy path (MEP) between an isolated skyrmion and the FM background for FGT-Li at $B_z = 1$ T (Fig. \ref{GNEB}(b)). From the MEP, we find the annihilation energy barrier, $\Delta E^{\text{Sk} \rightarrow \text{FM}}$, protecting the skyrmion from collapsing into the FM state, which is  
defined by the energy difference between the skyrmion state and the saddle point (SP). On the other hand, the nucleation barrier, $\Delta E^{\text{FM} \rightarrow \text{Sk}}$, is obtained from the energy difference between the SP and the FM state. 
The energy barrier $\Delta E^{\text{Sk} \rightarrow \text{FM}}$ arises due to the competition of exchange, MAE, and Zeeman energies which favor the FM state, and the DMI which protects the skyrmion from collapsing. The GNEB calculations also provide information about the mechanisms of magnetic transitions. From the analysis of spin structures near the SP, we find the skyrmion is annihilated via the radial symmetric collapse mechanism in which the skyrmion shrinks symmetrically to SP and then collapses into the FM state~\cite{muckel2021experimental}. The SP is also confirmed by the topological charge, calculated by $Q=\frac{1}{4\pi} \int_{\mathbb{R}^2} \V{m} \cdot (\frac{\partial \V{m}}{\partial x} \times \frac{\partial \V{m}}{\partial y})dxdy$, changes from $-1$ to $0$ at the SP, indicating that the SP coincides with the Bloch point.

We have calculated the energy barriers for skyrmion collapse
and annihilation as a function of magnetic field
(Fig.~\ref{Lifetime}(a)). The dots denote GNEB calculations based on parameters from DFT, while dashed lines are fitted curves. To avoid finite-size effects, we have performed GNEB calculations using $150 \times 150$ spin lattices for $B_z = 0.4 \sim 0.6$ T while a $100 \times 100$ spin lattice is used for $B_z > 0.6$ T. For $B_z < 0.4$ T, a very large lattice of more than $300 \times 300$ spins is required due to the large skyrmion size. However, this is computationally too demanding for the GNEB method. The skyrmion annihilation barrier is $\sim$ 300 meV at $B_z = 0.4$ T due to the relatively large skyrmion size (diameter $\approx$ 100 nm). Such a large skyrmion barrier is almost one order of magnitude higher than in previous work on skyrmion stability in 2D magnets with MAE = $K$/2, where $K = -0.86 \,\text{meV/uc}$ is the MAE estimated by DFT \cite{Dongzhe2022_fgt}. 

Upon increasing the external magnetic field, $\Delta E ^{\text{Sk} \rightarrow \text{FM}}$ decreases exponentially while $\Delta E ^{\text{FM} \rightarrow \text{Sk}}$ increases linearly. This indicates the nucleation becomes less favorable, and the probability for annihilation rises, overall shifting the stability towards the FM state. 

The stability of metastable magnetic skyrmions can be quantified by their mean lifetime, $\tau$, which is given by the Arrhenius law $\tau = f_{0}^{-1} \exp\!\left(\Delta E / (k_{\mathrm{B}} T)\right)$, where $\Delta E$, $f_{0}$, and $T$ are energy barrier, pre-exponential factor, and temperature, respectively \cite{bessarab2018lifetime,Malottki2019,muckel2021experimental,Dongzhe_PRB2024}. The calculated value of $f_{0}/T$ within harmonic transition state theory~\cite{bessarab2018lifetime} is shown in Fig.~\ref{Lifetime}(b)
(see SM~\cite{supplmat} for technical details). As expected, the pre-exponential factor depends on $B_z$. This effect is similar to that observed in ultrathin transition-metal films, which can be traced back to a change of entropy with skyrmion radius and profile~\cite{Malottki2019,varentcova2020toward}. 
We observe a non-monotonous progression of pre-exponential factor in the magnetic field regime $0.4-1.0~$T for both skyrmion collapse or creation transitions. We use Hessian eigenvalue spectra to analyze the origin of the non-monotonous progression and show that it originates from 
multiple crossings of eigenvalues above the magnon gap in presence of a skyrmion and therefore is a multi-magnon effect
(see Fig.~S6 in SM~\cite{supplmat} for details). 

From the obtained temperature and field dependence of the skyrmion lifetime (Fig.~\ref{Lifetime}(c)), we predict that isolated skyrmions in the FGT-Li monolayer are stable up to above one hour at a temperature of about 75K at about $B_z = 0.5$ T. Hence, these skyrmions can be probed, e.g., by scanning tunneling microscopy or Lorentz transmission electron microscopy experiments.

In summary, using a multiscale approach that combines \textit{ab initio} theory with atomistic spin simulations, we predict the formation of highly stable skyrmions down to nanometer size in lithium-decorated monolayer Fe$_3$GeTe$_2$. The strategy proposed in this work, lithium absorption on the surface of 2D magnets, is experimentally feasible using molecular-beam epitaxy or lithium-ion liquid regulation. In particular, we find very large skyrmion energy barriers in the FGT-Li monolayer, comparable to those observed in state-of-the-art transition-metal ultrathin films. We trace the origin to the competition between strong DMI, exchange frustration, and small MAE. We further demonstrate that nanoscale skyrmions are stable with lifetimes of one hour at temperatures up to 75 K. Our work provides a possible strategy to tailor magnetic interactions and topological magnetism in 2D magnets.

\section*{Acknowledgments}
	
This study has been supported through the ANR Grant No. ANR-22-CE24-0019. This work is supported by France 2030 government investment plan managed by the French National Research Agency under grant reference PEPR SPIN – [SPINTHEORY] ANR-22-EXSP-0009. S.~Ha and S.~He.~gratefully acknowledge financial support from the Deutsche Forschungsgemeinschaft (DFG, German Research Foundation) through SPP2137 ``Skyrmionics" (project no.~462602351). This work was performed using the HPC resources provided by CALMIP (Grant 2022/2025-[P21023]). 
	
	
\bibliography{References}

\end{document}